\documentstyle[aps,epsfig]{revtex}
\begin{document}
%\draft

\title{A microscopic description for the alpha decay of nuclei using a realistic effective interaction }

\author{D.N. Basu\thanks{E-mail:dnb@veccal.ernet.in}}
\address{Variable  Energy  Cyclotron  Centre,  1/AF Bidhan Nagar,
Kolkata 700 064, India}
\date{\today }
\maketitle
\begin{abstract}

      The process of $\alpha$ disintegration has been studied theoretically in the framework of a microscopic superasymmetric fission model (MSAFM). The nuclear interaction potential required for the $\alpha$ decay process has been calculated by  folding in the density distribution functions of the $\alpha$ nucleus and the daughter nucleus with a realistic effective interaction. The nuclear microscopic $\alpha$-nucleus potential thus obtained has been used along with the Coulomb interaction potential to calculate the action integral within the WKB approximation. This subsequently results in a parameter free calculation for the half lives of the $\alpha$ decays of nuclei. The model is successful in calculating the half lives of the $\alpha$ disintegration processes of nuclei.  

\end{abstract}


%\eject

      The first experimental observation of the $\alpha$ radioactivity \cite{r1,r2} was followed by a theoretical explanation in terms of quantum mechanical barrier penetration \cite{r3,r4} and was, at least, qualitatively successful. Since then the $\alpha$ decay half life measurements with constantly improving techniques and the quest for an explanation of their absolute magnitudes have continued unabated. Recently, predictions for $\alpha$ and various exotic decays have been made in detail by the analytical superasymmetric fission model (ASAFM) \cite{r5,r6} with reasonable success. This was followed by the cluster model (CM) calculations for  $\alpha$  decay \cite{r7} for even more comprehensive database with similar success. But  both the theoretical approaches described above use phenomenological potentials for the nuclear interactions. The ASAFM uses a parabolic potential approximation for the nuclear interaction potential, which is a rather unusual $\alpha$-nucleus interaction potential while the CM uses even stranger nuclear interaction potential  \cite{r7}. However, they provide a benchmark against which more microscopically based treatments can be judged.

      In the present work, the nuclear potentials have been obtained microscopically by double folding the $\alpha$ and daughter nuclei density distributions with realistic effective interaction. This procedure of obtaining nuclear interaction energy for the $\alpha$-nucleus interaction is absolutely fundamental in nature. Moreover, the use of a microscopic nuclear potential, free from adjustable parameters, for a wide range of $\alpha$-nucleus interaction is also theoretically a very radical approach. The double folding potential has then been utilised within a superasymmetric fission model description. This microscopic superasymmetric fission model (MSAFM) has been used to provide estimates of $\alpha$ decay half lives.  

      The microscopic nuclear potentials $V_N(R)$ have been obtained by double folding in the densities of the fragments $\alpha$ and daughter nuclei with the finite range realistic M3Y effective interacion as

\begin{equation}
 V_N(R) = \int \int \rho_1(\vec{r_1}) \rho_2(\vec{r_2}) v[|\vec{r_2} - \vec{r_1} + \vec{R}|] d^3r_1 d^3r_2 
\label{seqn1}
\end{equation}
\noindent
where the density distribution function $\rho_1$ for the $\alpha$ particle has the Gaussian form

\begin{equation}
 \rho_1(r) = 0.4229 exp( - 0.7024 r^2)
\label{seqn2}
\end{equation}                                                                                                                                           \noindent     
whose volume integral is equal to $A_\alpha ( = 4 )$, the mass number of $\alpha$-particle. The density distribution function $\rho_2$ used for the residual cluster, the daughter nucleus, has been chosen to be of the spherically symmetric form given by

\begin{equation}
 \rho_2(r) = \rho_0 / [ 1 + exp( (r-c) / a ) ]
\label{seqn3}
\end{equation}                                                                                                                                           \noindent     
where                        
 
\begin{equation}
 c = r_\rho ( 1 - \pi^2 a^2 / 3 r_\rho^2 ), ~~    r_\rho = 1.13 A_d^{1/3}  ~~   and ~~    a = 0.54 ~ fm
\label{seqn4}
\end{equation}
\noindent
and the value of $\rho_0$ is fixed by equating the volume integral of the density distribution function to the mass number $A_d$ of the residual daughter nucleus. The finite range M3Y effective interaction $v(s)$ appearing in the eqn.(1) is given by \cite{r8} 

\begin{equation}
 v(s) = 7999. \exp( - 4.s) / (4.s) - 2134. \exp( - 2.5s) / (2.5s)
\label{seqn5}
\end{equation}   
\noindent
The total interaction energy $E(R)$ between the $\alpha$ nucleus and the residual daughter nucleus is equal to the sum of the nuclear interaction energy, the Coulomb interaction energy and the centrifugal barrier. Thus

\begin{equation}
 E(R) = V_N(R) + V_C(R) + \hbar^2 l(l+1) / (2\mu R^2)
\label{seqn6}
\end{equation}   
\noindent
where $\mu = mA_\alpha A_d/A$  is the reduced mass, $A$ is the mass number of the parent nucleus and m is the nucleon mass measured in the units of $MeV/c^2$. The Coulomb potential $V_C(R)$ has been taken to be 

\begin{eqnarray}
 V_C(R) =&&Z_\alpha Z_d e^2/ R~~~~~~~~~~~~~~~~~~~~~~~~~~~~~~~~~~for~~~~R \geq R_c \nonumber\\
            =&&(Z_\alpha Z_d e^2/ 2R_c).[ 3 - (R/R_c)^2]~~~~~~~~~~for~~~~R\leq R_c 
\label{seqn7}
\end{eqnarray}   
\noindent
where $Z_\alpha$ and $Z_d$ are the atomic numbers of the $\alpha$-particle and the daughter nucleus respectively. The touching radial separation $R_c$ between the $\alpha$-particle and the daughter nucleus is given by $R_c = c_\alpha+c_d$ where $c_\alpha$ and $c_d$ has been obtained using eqn.(4). The energetics allow spontaneous emission of $\alpha$-particle only if the released energy 

\begin{equation}
 Q = M - ( M_\alpha + M_d)
\label{seqn8}
\end{equation}
\noindent
is a positive quantity, where $M$, $M_\alpha$ and $M_d$ are the atomic masses of the parent nucleus, the emitted $\alpha$-particle and the residual daughter nucleus, respectively,  expressed in the units of energy. It is important to mention here that the correctness of predictions for possible decay modes, therefore, rests on the accuracy of the ground state masses of nuclei.

      In the present model (MSAFM), the half life of the parent nucleus against the split into an $\alpha$ and a daughter is calculated using the WKB barrier penetration probability. The assault frequecy $\nu$ is obtained from the zero point vibration energy $E_v = (1/2)\hbar\omega = (1/2)h\nu$. The half life $T$ of the parent nucleus $(A, Z)$ against its split into an $\alpha$ $(A_\alpha, Z_\alpha)$ and a daughter $(A_d, Z_d)$  is given by

\begin{equation}
 T = [(h \ln2) / (2 E_v)] [1 + \exp(K)]
\label{seqn9}
\end{equation}
\noindent
where the action integral $K$ within the WKB approximation is given by

\begin{equation}
 K = (2/\hbar) \int_{R_a}^{R_b} {[2\mu (E(R) - E_v - Q)]}^{1/2} dR
\label{seqn10}
\end{equation}
\noindent
where $R_a$ and $R_b$ are the two turning points of the WKB action integral determined from the equations

\begin{equation}
 E(R_a)  = Q + E_v =  E(R_b)
\label{seqn11}
\end{equation} 
\noindent
It is important to mention here that the cluster model calculations \cite{r9} calculate assault frequency from the Q value in an oversimplistic manner. For the present calculations, from a fit to a selected set of experimental data on $\alpha$ emitters the following law \cite{r10} was found for the zero point vibration energies for the $\alpha$ decays
  
\begin{eqnarray}
 E_v =&& 0.1045.Q~~~~~~~~~~for~~even(Z)-even(N)~~parent~nuclei \nonumber\\
        =&& 0.0962.Q~~~~~~~~~~for~~odd(Z)-even(N)~~parent~nuclei  \nonumber\\
        =&& 0.0907.Q~~~~~~~~~~for~~even(Z)-odd(N)~~parent~nuclei \nonumber\\
        =&&0.0767.Q~~~~~~~~~~for~~odd(Z)-odd(N)~~parent~nuclei 
\label{seqn12}
\end{eqnarray} 
\noindent
which includes the shell and pairing effects. The values of the proportionality constants of $E_v$ with $Q$ is the largest for even-even parent and the smallest for the odd-odd one. If all other conditions are the same one may observe that with greater value of $E_v$, the life time is shortened indicating higher emission rate. The shell effects of $\alpha$ radioactivity is implicitly contained in the zero point vibration energy due to its proportionality with the Q value, which is maximum when the daughter nucleus has a magic number of neutrons and protons. 

      The two turning points of the action integral given by eqn.(10) have been obtained by solving eqns.(11) using microscopic double folding potential given by eqn.(1) along with the Coulomb potential given by eqn.(7) and the centrifugal barrier. Then the WKB action integral between these two turning points has been evaluated numerically using eqn.(1), eqn.(6), eqn.(7), eqn.(8) and eqn.(12). Finally, the half lives of the $\alpha$ decays have been calculated using eqn.(9) and eqn.(12).  

\begin{figure}[h]
\eject\centerline{\epsfig{file=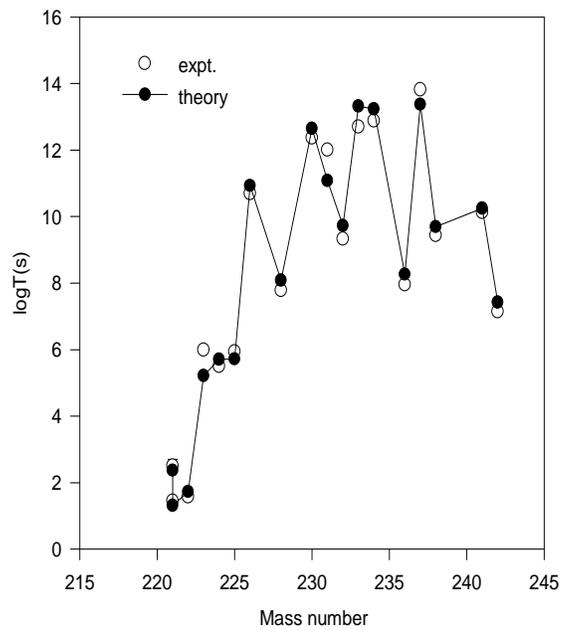,height=15cm,width=10cm}}
\caption
{Logarithmic half lives for $\alpha$ decays plotted against the parent mass number. The experimental data are shown by the open circles. The solid circles are the results of present (MSAFM) calculations for the corresponding experimental data. The continuous line connects these calculated values. }
\label{fig1}
\end{figure}

      In the present illustrative calculations the same set of experimental data of reference \cite{r11} for the $\alpha$ decay half lives have been chosen for comparison with the present theoretical calculation for which the experimental ground state masses for the parent and daughter nuclei are available. This set was selected because there is no uncertainty in the determination of the released energy Q (given by eqn.(8)) which is one of the crucial quantity for quantitative prediction of decay half lives. A normalisation factor of 0.9 for the microscopic nuclear potential has been used to obtain the optimum fit. In Fig.~\ref{fig1} the experimental data for logarithmic $\alpha$ decay half lives \cite{r11} have been plotted against the mass numbers of parent nuclei along with the results of the present calculations for zero angular momentum of the fragments. In the figure the open circles depict the experimental data while the continuous line with solid circles represent the corresponding calculations of the present model (MSAFM). The decay modes and the experimental values for their half lives have been presented in Table 1. The corresponding results of the present calculations of superasymmetric fission model with microscopic potentials (MSAFM) are also presented along with the results of the cluster model (CM) calculations of 1993 \cite{r7} so as to facilitate the comparison of the results of the phenomenological calculations with the present one. The exact Q values used by the present calculation using the experimental ground state masses have also been presented along with those obtained by eqn.(4) of \cite{r7}.
 
\begin{table}
\caption{Comparison between Measured and Calculated $\alpha$ decay Half-Lives}
\begin{tabular}{ccccccccccc}
Parent &       &Daughter &      &CM       & CM    &MSAFM & MSAFM  &  Expt. & Error &      \\ 
Z  & A & $Z_d$  & $A_d$  & Q &  logT(s) & Q & logT(s) &  logT(s) &  $\pm$err \\ \hline

  87. &221. & 85.& 217.&  6.273   & 3.23 &  6.47&    2.37          &  2.50     & .20  \\ 
  88. &221. & 86. &217. & 6.764   & 1.62 &  6.89  &  1.31    &        1.45     & .03  \\ 
  88. &222. & 86. &218.  &6.710   & 1.61  & 6.68    &1.73      &      1.58      &.05  \\ 
  88. &223. & 86. &219.  &  -----   &  -----   & 5.98&    5.21        &    5.995     &.001  \\ 
  88. &224.  &86. &220. & 5.823    &5.57  & 5.79  &  5.70          &  5.50      &.04  \\ 
  89. &225.  &87. &221.  &5.417   & 8.43 &  5.94    &5.72  &          5.94     & .04   \\
  88. &226.  &86. &222. & 4.904  & 10.79  & 4.87&   10.94  &         10.703   &  .002 \\  
  90. &228.  &88. &224. & 5.555   & 7.95&   5.53   & 8.08     &       7.781    & .001   \\
  91. &231.  &89. &227. & 4.853   &13.00 &  5.15&   11.08     &      12.014   &  .001 \\  
  90. &230.  &88. &226. & 4.806  & 12.56   &4.78  & 12.65       &    12.376    & .001  \\ 
  92. &232.  &90. &228. & 5.450   & 9.56&   5.42    &9.73          &  9.337    & .001   \\
  92. &233.  &90. &229. & 4.945  & 12.93 &  4.92&   13.32 &          12.701   &  .005  \\ 
  92. &234.  &90. &230. & 4.895  & 13.04   &4.86  & 13.24   &        12.889    & .001\\   
  94. &236.  &92. &232. & 5.904  &  8.04&   5.87    &8.27      &      7.954     &.001  \\ 
  93. &237.  &91. &233. & 4.756  & 14.81 &  4.96&   13.38      &     13.829    & .002  \\ 
  94. &238.  &92. &234. & 5.631  &  9.54   &5.60   & 9.70         &   9.4423    &.0004  \\ 
  95. &241.  &93. &237. & 5.616  & 10.30 &  5.64   &10.25         &  10.135    & .0005  \\ 
  96. &242.  &94. &238. & 6.254  &  7.23   &6.22 &   7.42            &7.1485  &  .0002  \\

\end{tabular} 
\end{table}

      The results of the present calculations of the MSAFM have been found to predict the general trend  of experimental data very well. The quantitave agreement with experimental data is excellent. The degree of reliability of the MSAFM predictions for $\alpha$ decay lifetimes are comparable to the CM \cite{r7} predictions and are better than the liquid drop description \cite{r11} for most of the cases. The recent result of the preformed cluster model \cite{r9} calculation for the $\alpha$ decay half-life of $^{242} Cm$ is also much worse compared to that obtained from present calculations. The exact Q values obtained from the ground state masses and used in the present calculations differ slightly from those obtained from the measured $\alpha$ particle kinetic energy by applying a standard recoil correction, as well as an electron shielding correction in a systematic manner and used in the CM \cite{r7} calculations.
                                 
      The half lives for $\alpha$-radioactivity have been analyzed with microscopic nuclear potentials which are based on profound theoretical basis. The results of the present calculations with MSAFM are in good agreement over a wide range of experimental data and are comparable to the best available theoretical calculations \cite{r7} of CM which use unrealistic nuclear interaction potentials that do not have any microscopic basis. Refinements such as introduction of dissipation while tunneling through the barrier or incorporating the dynamic shape deformations in the density distributions of the clusters may further improve results. Corrections in the Coulomb interaction energy in the overlapping region for the deformed nuclear system may also be incorporated. It is worthwhile to mention that as the first illustrative calculations using realistic microscopic nuclear interaction potentials, the results obtained for the $\alpha$ radioactive decay lifetimes are noteworthy. Such calculations may be extended to provide reasonable estimates of the lifetimes of nuclear decays by $\alpha$ emissions for the entire domain of exotic nuclei provided that good estimates of the Q values are available.

\end{document}